\documentclass[showpacs]{revtex4}
\usepackage{indentfirst}
\begin{document}
\title{\bf Twelve-quark hypernuclei with $A=4$ in relativistic quark-gluon model}
\author{S.M. Gerasyuta}
\email{gerasyuta@SG6488.spb.edu}
\author{E.E. Matskevich}
\email{matskev@pobox.spbu.ru}
\affiliation{Department of Physics, St. Petersburg State Forest Technical
University, Institutski Per. 5, St. Petersburg 194021, Russia}
\begin{abstract}
Hypernuclei $ ^4_Y He$, $ ^4_Y H$, $ ^4_{YY} He$, $ ^4_{YY} H$, where
$Y=\Lambda$, $\Sigma_0$,  $\Sigma_+$, $\Sigma_-$, $A=4$ are considered
using the relativistic twelve-quark equations in the framework of the
dispersion relation technique. Hypernuclei as the systems of interacting
quarks and gluons are considered. The relativistic twelve-quark amplitudes
of hypernuclei, including $u$, $d$, $s$ quarks are constructed. The approximate
solutions of these equations are obtained using a method based on the
extraction of leading singularities of the amplitudes. The poles of the
multiquark amplitudes allow us to determine the masses of hypernuclei
with the atomic (baryon) number $A=B=4$. The mass of state $ ^4_{\Lambda}He$
with the isospin projection $I_3=\frac{1}{2}$ and the spin-parity $J^P=0^+$ is
equal to $M=3922\, MeV$. The mass of $ ^4_{\Lambda\Lambda}H$ $M=4118\, MeV$
with the isospin projection $I_3=0$ and the spin-parity $J^P=0^+$ is
calculated. We predict the mass spectrum of hypernuclei with $A=4$, which
is valuable to further experimental study of the hypernuclei.
\end{abstract}
\maketitle
\section{Introduction.}
The hypernuclear physics experimental program, and the difficulties
arised in accurately determining rates for low-energy nuclear reactions,
warrant continued effort in the application of LQCD to nuclear physics
 \cite{1, 2, 3, 4, 5, 6, 7, 8, 9, 10, 11, 12}.

In the limit of flavor $SU(3)$ symmetry at the physical strange quark mass
with quantum chromodynamics (without electromagnetic interactions) the
binding energies of a range of nuclei and hypernuclei with atomic number
$A\le 4$ and strangeness $|S|\le 2$, including the deuteron, H-dibaryon,
$ ^3 He$, $ ^3_{\Lambda} He$, $ ^4 He$, $ ^4_{\Lambda} He$ and $ ^4_{\Lambda\Lambda} He$
are calculated. From lattice QCD calculations performed with $n_f=3$ dynamical
light quark using an isotropic discretization, the nuclear states are
extracted \cite{13, 14, 15}. It is now clear that the spectrum of nuclei
and hypernuclei changes dramatically from light quark masses.

In our recent paper \cite{16} the relativistic six-quark equations are found
in the framework of coupled-channel formalism. The dynamical mixing between
the subamplitudes of hexaquark is considered. The six-quark amplitudes
of dibaryons are calculated. The poles of these amplitudes determine the
masses of dibaryons. We calculated the contribution of six-quark
subamplitudes to the hexaquark amplitudes. The model in question has only three
parameters: the cutoff parameter $\Lambda=11$ and gluon coupling constants $g_0$
and $g_1$. These parameters are determined by the $\Lambda\Lambda$ and di-$\Omega$
masses. In our model the correlation of gluon coupling constants $g_0$ and $g_1$
is similar to the $S$-wave baryon ones \cite{17}.

In the previous paper \cite{18}, $ ^3 He$ is considered. The relativistic
nine-quark equations are derived in the framework of the dispersion relation
technique. The dynamical mixing between the subamplitudes of $ ^3 He$ is taken
into account. The relativistic nine-quark amplitudes of $ ^3 He$, including the
$u$, $d$ quarks are calculated. The approximate solutions of these equations
were obtained using a method based on the extraction of leading singularities
of the amplitudes. The pole of the nonaquark amplitudes determined the mass of $ ^3 He$.

The experimental mass value of $ ^3 He$ is equal to $M=2808.39\, MeV$.
The experimental data of the hypertriton mass is $M=2991.17\, MeV$.
This model use only three parameters, which are determined by the following masses:
the cutoff $\Lambda=9.0$ and the gluon coupling constant $g=0.2122$.
The mass of the $u$-quark is $m=410\, MeV$, and the mass of strange quark $m_s=607\, MeV$,
which takes into account the confinement potential (the shift mass is equal to $50\, MeV$)
\cite{19}.

\begin{table}
\caption{Masses and binding energies of $ ^4 He$. Parameters of model: $\Lambda=6.355$,
$g=0.2122$, $m=410\, MeV$, $m_s=594\, MeV$.}\label{tab1}
\begin{tabular}{|ll|c|c|c|c|c|c|c|}
\hline
\multicolumn{2}{|c|}{hypernuclei} & quark content & $Q$ & $I_3$ & $I$ & $J^P$ & mass, MeV & binding energy, MeV \\ [5pt]
\hline
$ ^4_{\Lambda} He$  &  ($ppn\Lambda$)  & $uud\, uud\, udd \, uds$ & 2 & $\frac{1}{2}$  & $\frac{1}{2}$, $\frac{3}{2}$                & $0^+$, $1^+$ & 3922 & 10  \\ [5pt]
$ ^4_{\Sigma^0} He$ &  ($ppn\Sigma^0$) & $uud\, uud\, udd \, uds$ & 2 & $\frac{1}{2}$  & $\frac{1}{2}$, $\frac{3}{2}$, $\frac{5}{2}$ & $0^+$, $1^+$ & 3922 & 87  \\ [5pt]
$ ^4_{\Sigma^+} He$ &  ($ppn\Sigma^+$) & $uud\, uud\, udd \, uus$ & 3 & $\frac{3}{2}$  & $\frac{3}{2}$, $\frac{5}{2}$                & $0^+$, $1^+$ & 3904 & 101 \\ [5pt]
$ ^4_{\Sigma^-} He$ &  ($ppn\Sigma^-$) & $uud\, uud\, udd \, dds$ & 1 & $-\frac{1}{2}$ & $\frac{1}{2}$, $\frac{3}{2}$, $\frac{5}{2}$ & $0^+$, $1^+$ & 3933 & 81  \\ [5pt]
$ ^4_{\Xi^0} He$    &  ($ppn\Xi^0$)    & $uud\, uud\, udd \, uss$ & 2 & 1              & 1, 2                                        & $0^+$, $1^+$ & 4106 & 25  \\ [5pt]
$ ^4_{\Xi^-} He$    &  ($ppn\Xi^-$)    & $uud\, uud\, udd \, dss$ & 1 & 0              & 0, 1, 2                                     & $0^+$, $1^+$ & 4121 & 17  \\ [5pt]
$ ^4_{\Omega} He$   &  ($ppn\Omega$)   & $uud\, uud\, udd \, sss$ & 1 & $\frac{1}{2}$  & $\frac{1}{2}$, $\frac{3}{2}$                & $1^+$, $2^+$ & 4293 & 196 \\ [5pt]
\hline
\end{tabular}
\end{table}

\begin{table}
\caption{Masses and binding energies of $ ^4 H$ hypernuclei.}\label{tab2}
\begin{tabular}{|ll|c|c|c|c|c|c|c|}
\hline
\multicolumn{2}{|c|}{hypernuclei} & quark content & $Q$ & $I_3$ & $I$ & $J^P$ & mass, MeV & binding energy, MeV \\ [5pt]
\hline
$ ^4_{\Lambda} H$  & ($pnn\Lambda$)  & $uud\, udd\, udd \, uds$ & 1 & $-\frac{1}{2}$ & $\frac{1}{2}$, $\frac{3}{2}$                & $0^+$, $1^+$ & 3922 & 12  \\ [5pt]
$ ^4_{\Sigma^0} H$ & ($pnn\Sigma^0$) & $uud\, udd\, udd \, uds$ & 1 & $-\frac{1}{2}$ & $\frac{1}{2}$, $\frac{3}{2}$, $\frac{5}{2}$ & $0^+$, $1^+$ & 3922 & 89  \\ [5pt]
$ ^4_{\Sigma^-} H$ & ($pnn\Sigma^-$) & $uud\, udd\, udd \, dds$ & 0 & $-\frac{3}{2}$ & $\frac{3}{2}$, $\frac{5}{2}$                & $0^+$, $1^+$ & 3904 & 112 \\ [5pt]
$ ^4_{\Sigma^+} H$ & ($pnn\Sigma^+$) & $uud\, udd\, udd \, uus$ & 2 & $\frac{1}{2}$  & $\frac{1}{2}$, $\frac{3}{2}$, $\frac{5}{2}$ & $0^+$, $1^+$ & 3933 & 74  \\ [5pt]
$ ^4_{\Xi^-} H$    & ($pnn\Xi^-$)    & $uud\, udd\, udd \, dss$ & 0 & -1             & 1, 2                                        & $0^+$, $1^+$ & 4106 & 34  \\ [5pt]
$ ^4_{\Xi^0} H$    & ($pnn\Xi^0$)    & $uud\, udd\, udd \, uss$ & 1 & 0              & 0, 1, 2                                     & $0^+$, $1^+$ & 4121 & 12  \\ [5pt]
$ ^4_{\Omega} H$   & ($pnn\Omega$)   & $uud\, udd\, udd \, sss$ & 0 & $-\frac{1}{2}$ & $\frac{1}{2}$, $\frac{3}{2}$                & $1^+$, $2^+$ & 4293 & 198 \\ [5pt]
\hline
\end{tabular}
\end{table}

The relativistic nona-amplitudes of low-lying hypernuclei,
including the three flavors ($u$, $d$, $s$) are calculated. The degeneracy of the
isospin 0, 1, 2 is predicted in the lowest hypernuclei. It is the property of our
approach. The low-lying hypernuclei with the spin-parity $J^P=\frac{1}{2}^+$,
$\frac{3}{2}^+$ are calculated. We calculated the masses and the binding energies
of 12 hypernuclei with $A=3$. The binding energy is small for the states
$ ^3_{\Lambda} He$, $ ^3_{\Lambda} H$, and $nn\Lambda$. In the other cases, the
binding energies are large, $\sim$ 50 -- 100 $MeV$. We have calculated only five
systems of equations; therefore, the masses of hypernuclei are degenerated. We do
not include the electromagnetic effect contribution.

Hypernuclei spectroscopy is enjoying an experimental renaissance with
ongoing and planned program at DA$\Phi$NE, FAIR, Jefferson Lab, J-PARC
and Mainz providing motivation for enhanced theoretical efforts
(for a recent review, see Ref. \cite{20}).

In the chiral soliton approach using the bound state rigid oscillator
version of the $SU(3)$ quantization model \cite{21, 22} the binding
energies of neutron-rich are estimated.

Recently, a new direction of such studies, the studies of neutron-rich
hypernuclei with strangeness $S=-1$ got new impact due to discovery
of the hypernucleus $ ^6_{\Lambda}H$ (heavy hyperhydrogen) by the FINUDA
Collaboration \cite{23}.

\begin{table}
\caption{Masses and binding energies of $ ^4_{YY} He$.}\label{tab3}
\begin{tabular}{|ll|c|c|c|c|c|c|c|}
\hline
\multicolumn{2}{|c|}{hypernuclei} & quark content & $Q$ & $I_3$ & $I$ & $J^P$ & mass, MeV & binding energy, MeV \\ [5pt]
\hline
$ ^4_{\Lambda\Lambda} He$   & ($pp\Lambda\Lambda$)    & $uud\, uud\, uds \, uds$ & 2 & 1  & 1       & $0^+$        & 4127 & -19 \\ [5pt]
$ ^4_{\Lambda\Sigma^0} He$  & ($pp\Lambda\Sigma^0$)   & $uud\, uud\, uds \, uds$ & 2 & 1  & 1, 2    & $0^+$, $1^+$ & 4127 & 58  \\ [5pt]
$ ^4_{\Sigma^0\Sigma^0} He$ & ($pp\Sigma^0\Sigma^0$)  & $uud\, uud\, uds \, uds$ & 2 & 1  & 1, 2, 3 & $0^+$        & 4127 & 135 \\ [5pt]
$ ^4_{\Sigma^+\Sigma^+} He$ & ($pp\Sigma^+\Sigma^+$)  & $uud\, uud\, uus \, uus$ & 4 & 3  & 3       & $0^+$        & 4008 & 246 \\ [5pt]
$ ^4_{\Sigma^-\Sigma^-} He$ & ($pp\Sigma^-\Sigma^-$)  & $uud\, uud\, dds \, dds$ & 0 & -1 & 1, 2, 3 & $0^+$        & 4150 & 122 \\ [5pt]
$ ^4_{\Lambda\Sigma^+} He$  & ($pp\Lambda\Sigma^+$)   & $uud\, uud\, uds \, uus$ & 3 & 2  & 2       & $0^+$, $1^+$ & 4086 & 95  \\ [5pt]
$ ^4_{\Lambda\Sigma^-} He$  & ($pp\Lambda\Sigma^-$)   & $uud\, uud\, uds \, dds$ & 1 & 0  & 0, 1, 2 & $0^+$, $1^+$ & 4114 & 76  \\ [5pt]
$ ^4_{\Sigma^+\Sigma^-} He$ & ($pp\Sigma^+\Sigma^-$ ) & $uud\, uud\, uus \, dds$ & 2 & 1  & 1, 2, 3 & $0^+$, $1^+$ & 4105 & 158 \\ [5pt]
\hline
\end{tabular}
\end{table}

\begin{table}
\caption{Masses and binding energies of $ ^4_{YY} H$.}\label{tab4}
\begin{tabular}{|ll|c|c|c|c|c|c|c|}
\hline
\multicolumn{2}{|c|}{hypernuclei} & quark content & $Q$ & $I_3$ & $I$ & $J^P$ & mass, MeV & binding energy, MeV \\ [5pt]
\hline
$ ^4_{\Lambda\Lambda} H$   & ($pn\Lambda\Lambda$)   & $uud\, udd\, uds \, uds$ & 1  & 0  & 0, 1       & $0^+$, $1^+$        & 4118 & -8  \\ [5pt]
$ ^4_{\Lambda\Sigma^0} H$  & ($pn\Lambda\Sigma^0$)  & $uud\, udd\, uds \, uds$ & 1  & 0  & 0, 1, 2    & $0^+$, $1^+$, $2^+$ & 4118 & 69  \\ [5pt]
$ ^4_{\Sigma^0\Sigma^0} H$ & ($pn\Sigma^0\Sigma^0$) & $uud\, udd\, uds \, uds$ & 1  & 0  & 0, 1, 2, 3 & $0^+$, $1^+$        & 4118 & 146 \\ [5pt]
$ ^4_{\Sigma^+\Sigma^+} H$ & ($pn\Sigma^+\Sigma^+$) & $uud\, udd\, uus \, uus$ & 3  & 2  & 2, 3       & $0^+$, $1^+$        & 4081 & 175 \\ [5pt]
$ ^4_{\Sigma^-\Sigma^-} H$ & ($pn\Sigma^-\Sigma^-$) & $uud\, udd\, dds \, dds$ & -1 & -2 & 2, 3       & $0^+$, $1^+$        & 4081 & 193 \\ [5pt]
$ ^4_{\Lambda\Sigma^+} H$  & ($pn\Lambda\Sigma^+$)  & $uud\, udd\, uds \, uus$ & 2  & 1  & 1, 2       & $0^+$, $1^+$, $2^+$ & 4099 & 84  \\ [5pt]
$ ^4_{\Lambda\Sigma^-} H$  & ($pn\Lambda\Sigma^-$)  & $uud\, udd\, uds \, dds$ & 0  & -1 & 1, 2       & $0^+$, $1^+$, $2^+$ & 4099 & 93  \\ [5pt]
$ ^4_{\Sigma^+\Sigma^-} H$ & ($pn\Sigma^+\Sigma^-$) & $uud\, udd\, uus \, dds$ & 1  & 0  & 0, 1, 2, 3 & $0^+$, $1^+$, $2^+$ & 4109 & 156 \\ [5pt]
\hline
\end{tabular}
\end{table}

In the present paper, the hypernuclei $ ^4_Y He$, $ ^4_Y H$, $ ^4_{YY} He$, $ ^4_{YY} H$, where
$Y=\Lambda$, $\Sigma_0$,  $\Sigma_+$, $\Sigma_-$; $A=4$ are considered.
The approximate solutions of the twelve-quark equations are obtained using
the method based on the extraction of leading singularities of the amplitude.
The poles of these amplitudes determine the masses of the hypernuclei.

In Sec. II, the relativistic twelve-quark amplitudes, including the three
flavors ($u$, $d$, $s$) are constructed. We derived the dynamical mixing
between the subamplitudes of the hypernuclei. Sec. III is devoted to the calculation
results for the masses and binding energies of $ ^4 He$, $ ^4 H$ and $nnYY$
hypernuclei (Tables \ref{tab1} -- \ref{tab5}).

In conclusion, the status of the considered model is discussed (Table \ref{tab6}).

\begin{table}
\caption{Masses and binding energies of $nnYY$.}\label{tab5}
\begin{tabular}{|l|c|c|c|c|c|c|c|}
\hline
hypernuclei & quark content & $Q$ & $I_3$ & $I$ & $J^P$ & mass, MeV & binding energy, MeV \\ [5pt]
\hline
$nn\Lambda\Lambda$   & $udd\, udd\, uds \, uds$ & 0  & -1 & 1       & $0^+$        & 4127 & -15 \\ [5pt]
$nn\Lambda\Sigma^0$  & $udd\, udd\, uds \, uds$ & 0  & -1 & 1, 2    & $0^+$, $1^+$ & 4127 & 62  \\ [5pt]
$nn\Sigma^0\Sigma^0$ & $udd\, udd\, uds \, uds$ & 0  & -1 & 1, 2, 3 & $0^+$        & 4127 & 139 \\ [5pt]
$nn\Sigma^+\Sigma^+$ & $udd\, udd\, uus \, uus$ & 2  & 1  & 1, 2, 3 & $0^+$        & 4150 & 108 \\ [5pt]
$nn\Sigma^-\Sigma^-$ & $udd\, udd\, dds \, dds$ & -2 & -3 & 3       & $0^+$        & 4008 & 268 \\ [5pt]
$nn\Lambda\Sigma^+$  & $udd\, udd\, uds \, uus$ & 1  & 0  & 0, 1, 2 & $0^+$, $1^+$ & 4114 & 71  \\ [5pt]
$nn\Lambda\Sigma^-$  & $udd\, udd\, uds \, dds$ & -1 & -2 & 2       & $0^+$, $1^+$ & 4086 & 108 \\ [5pt]
$nn\Sigma^+\Sigma^-$ & $udd\, udd\, uus \, dds$ & 0  & -1 & 1, 2, 3 & $0^+$, $1^+$ & 4105 & 162 \\ [5pt]
\hline
\end{tabular}
\end{table}

\begin{table}
\caption{Binding energies of low-lying nuclei and hypernuclei (NPLQCD Collaboration).}\label{tab6}
\begin{tabular}{|c|c|c|c|c|c|}
\hline
State & $A$ & $S$ & $I$ & $J^P$ & binding energy ($B^{\infty}$), MeV \\ [5pt]
\hline
$ ^4_2 He$                & 4 &  0 & 0             & $0^+$ & 107 \\ [5pt]
$ ^4_{\Lambda} He$        & 4 & -1 & $\frac{1}{2}$ & $0^+$ & 107 \\ [5pt]
$ ^4_{\Lambda} H$         & 4 & -1 & $\frac{1}{2}$ & $0^+$ & 107 \\ [5pt]
$ ^4_{\Lambda\Lambda} He$ & 4 & -2 & 1             & $0^+$ & 156 \\ [5pt]
$ ^4_{\Lambda\Lambda} H$  & 4 & -2 & 0             & $0^+$ & 156 \\ [5pt]
$nn\Lambda\Lambda$        & 4 & -2 & 1             & $0^+$ & 156 \\ [5pt]
\hline
\end{tabular}
\end{table}

In the Appendix A, the coefficients of the coupled equations reduced amplitude
$\alpha_4$ ($I_3=\frac{1}{2}$, $J^P=0^+$ $ ^4_{\Lambda} He$) are given.

The Appendix B shown the graphical equations of the reduced amplitude
$\alpha_4^{1^{uu}1^{uu}1^{dd}0^{ds}}$ ($I_3=\frac{1}{2}$, $J^P=0^+$ $ ^4_{\Lambda} He$).

The interaction of quarks and gluons is described with the functions $I_1$ -- $I_{46}$ in Appendix C.

\section{Twelve-quark amplitudes of hypernuclei.}

We derive the relativistic twelve-quark equations in the framework of the
dispersion relation technique. The planar diagrams are used; the other
diagrams are neglected due to the rules of $1/N_c$ expansion \cite{24, 25, 26}.

The current generates a twelve-quark system. The correct equations for the
amplitude are obtained by taking into account all possible subamplitudes.
Then one should represent a twelve-particle amplitude as a sum of 66 subamplitudes:

\begin{eqnarray}
\label{1}
A=\sum\limits_{i<j \atop i, j=1}^{12} A_{ij}\, . \end{eqnarray}

This defines the division of the diagrams into groups according to the
certain pair interaction of particles. The total amplitude can be
represented graphically as a sum of diagrams. It allows us to consider only
one group of diagrams and the amplitude corresponding to them, for example
$A_{12}$. The relativistic generalization of the Faddeev-Yakubovsky
approach is used \cite{27, 28}. In our case, the hyperhydrogens and
hyperheliums consist of the similar equations and therefore, the masses
and other properties of these states are resembling. The quark content
of all states is given in the Tables \ref{tab1} -- \ref{tab5}.
The pairwise interaction of all twelve quarks is taken into account.
The set of diagrams associated with the amplitude $A_{12}$ can further be
broken down into some groups corresponding to subamplitudes.

In order to represent the subamplitudes $A_i$ in the form of a dispersion relation
it is necessary to define the amplitudes of quark-quark interactions $b_n(s_{ik})$.
The pair quarks amplitudes $qq\rightarrow qq$ are calculated in
the framework of the dispersion $N/D$ method with the input four-fermion
interaction with quantum numbers of the gluon \cite{29}.
We use the results of our relativistic quark model \cite{29} and write
down the pair quarks amplitudes in the form:

\begin{equation}
\label{2}
b_n(s_{ik})=\frac{G^2_n(s_{ik})}
{1-B_n(s_{ik})} \, ,\end{equation}

\begin{equation}
\label{3}
B_n(s_{ik})=\int\limits_{(m_i+m_k)^2}^{\Lambda_n}
\, \frac{ds'_{ik}}{\pi}\frac{\rho_n(s'_{ik})G^2_n(s'_{ik})}
{s'_{ik}-s_{ik}} \, .\end{equation}

\noindent
Here, $s_{ik}$ is the two-particle subenergy squared. $G_n(s_{ik})$ are
the quark-quark vertex functions (Table \ref{tab7}). $B_n(s_{ik})$, $\rho_n (s_{ik})$
are the Chew-Mandelstam functions with cutoff $\Lambda_n$ \cite{30} and
the phase space, respectively:

\begin{eqnarray}
\label{4}
\rho_n (s_{ik},J^P)&=&\left(\alpha(J^P,n) \frac{s_{ik}}{(m_i+m_k)^2}
+\beta(J^P,n)+\delta(J^P,n) \frac{(m_i-m_k)^2}{s_{ik}}\right)
\nonumber\\
&&\nonumber\\
&\times & \frac{\sqrt{(s_{ik}-(m_i+m_k)^2)(s_{ik}-(m_i-m_k)^2)}}
{s_{ik}}\, .
\end{eqnarray}

The coefficients $\alpha(J^P,n)$, $\beta(J^P,n)$ and
$\delta(J^P,n)$ are given in Table \ref{tab7}.

\begin{table}
\caption{The vertex functions and coefficients of Chew-Mandelstam functions.}\label{tab7}
\begin{tabular}{|c|c|c|c|c|c|}
\hline
\, $n$ \, & \, $J^P$ \, & $G_n^2(s_{kl})$ & \, $\alpha_n$ \, & $\beta_n$ & \, $\delta_n$ \, \\
\hline
& & & & & \\
1 & $0^+$ & $\frac{4g}{3}-\frac{8g m_{kl}^2}{(3s_{kl})}$
& $\frac{1}{2}$ & $-\frac{1}{2}\frac{(m_k-m_l)^2}{(m_k+m_l)^2}$ & $0$ \\
& & & & & \\
2 & $1^+$ & $\frac{2g}{3}$ & $\frac{1}{3}$
& $\frac{4m_k m_l}{3(m_k+m_l)^2}-\frac{1}{6}$
& $-\frac{1}{6}$ \\
& & & & & \\
\hline
\end{tabular}
\end{table}

Here $n=1$ coresponds to $qq$-pairs with $J^P=0^+$, $n=2$ describes
the $qq$ pairs with $J^P=1^+$.

In our model the interacting quarks do not produce a bound
states; therefore the integration is carried out from the threshold
$(m_i+m_k)^2$ to the cutoff $\Lambda_n$.

Let us extract singularities in the coupled equations and obtain
the reduced amplitudes $\alpha_i$.

At first, we obtain the system of 20 equations for the $ ^4_2 He$
with the isospin projection $I_3=0$, the spin-parity $J^P=0^+$ $(ppnn)$.
The reduced amplitudes $\alpha_1$ are determined by the channels:
$1^{uu}$, $0^{ud}$, $1^{dd}$.
The $\alpha_2$ are constructed as
$1^{uu}1^{uu}$, $1^{uu}0^{ud}$, $1^{uu}1^{dd}$, $0^{ud}0^{ud}$,
$0^{ud}1^{dd}$, $1^{dd}1^{dd}$.
The reduced amplitudes $\alpha_3$ are the following:

\begin{eqnarray}
\label{5}
1^{uu}1^{uu}0^{ud}, \, 1^{uu}1^{uu}1^{dd},
1^{uu}0^{ud}1^{dd}, \, 1^{uu}1^{dd}1^{dd}, \, 0^{ud}1^{dd}1^{dd}.
\end{eqnarray}

The $\alpha_4$ are just:

\begin{eqnarray}
\label{6}
&&1^{uu}1^{uu}0^{ud}0^{ud}, \, 1^{uu}1^{uu}0^{ud}1^{dd}, \, 1^{uu}1^{uu}1^{dd}1^{dd},
 \nonumber
\\
&&1^{uu}0^{ud}0^{ud}1^{dd}, \, 1^{uu}0^{ud}1^{dd}1^{dd}, \, 0^{ud}0^{ud}1^{dd}1^{dd}.
\end{eqnarray}

Here the $\alpha_1$ are determined by the diquarks, the $\alpha_2$
includes the two diquarks and eight quarks. $\alpha_3$ defines
the three diquarks and six quarks. The $\alpha_4$ allows us to
consider the $ppnn$ ($uud\, uud\, udd\, udd$) $ ^4_2 He$ state.

Then we calculate the solution of 33 equations for the hypernucleous $ ^4_{\Lambda} He$
($ppn\Lambda$) with the isospin projection $I_3=\frac{1}{2}$ and the spin-parity
$J^P=0^+$ (Table \ref{tab1}).

The reduced amplitudes $\alpha_1$, $\alpha_2$ with the one $s$-quark are
similar to:

\begin{eqnarray}
\label{7}
\alpha_1^{1^{uu}}, \, \alpha_1^{1^{dd}}, \, \alpha_1^{0^{ud}}, \, \alpha_1^{0^{us}}, \, \alpha_1^{0^{ds}};
\end{eqnarray}

\begin{eqnarray}
\label{8}
&&\alpha_2^{1^{uu}1^{uu}}, \, \alpha_2^{1^{uu}1^{dd}}, \, \alpha_2^{1^{uu}0^{ud}},
 \, \alpha_2^{1^{uu}0^{us}}, \, \alpha_2^{1^{uu}0^{ds}}, \, \alpha_2^{1^{dd}1^{dd}},
 \nonumber
\\
&&\alpha_2^{1^{dd}0^{ud}}, \, \alpha_2^{1^{dd}0^{us}}, \, \alpha_2^{1^{dd}0^{ds}},
 \, \alpha_2^{0^{ud}0^{ud}}, \, \alpha_2^{0^{ud}0^{us}}, \, \alpha_2^{0^{ud}0^{ds}}.
\end{eqnarray}

We have to add the reduced amplitudes $\alpha_3$, $\alpha_4$

\begin{eqnarray}
\label{9}
&&\alpha_3^{1^{uu}1^{uu}1^{dd}}, \, \alpha_3^{1^{uu}1^{uu}0^{ud}}, \, \alpha_3^{1^{uu}1^{uu}0^{us}}, \, \alpha_3^{1^{uu}1^{uu}0^{ds}},
 \nonumber
\\
&&\alpha_3^{1^{uu}1^{dd}0^{ud}}, \, \alpha_3^{1^{uu}1^{dd}0^{us}}, \, \alpha_3^{1^{uu}1^{dd}0^{ds}};
\end{eqnarray}

\begin{eqnarray}
\label{10}
&&\alpha_4^{1^{uu}1^{uu}1^{dd}0^{ud}}, \, \alpha_4^{1^{uu}1^{uu}1^{dd}0^{us}}, \, \alpha_4^{1^{uu}1^{uu}1^{dd}0^{ds}},
 \nonumber
\\
&&\alpha_4^{1^{uu}1^{uu}0^{ud}0^{ud}}, \, \alpha_4^{1^{uu}1^{uu}0^{ud}0^{us}}, \, \alpha_4^{1^{uu}1^{uu}0^{ud}0^{ds}},
 \nonumber
\\
&&\alpha_4^{1^{uu}1^{dd}0^{ud}0^{ud}}, \, \alpha_4^{1^{uu}1^{dd}0^{ud}0^{us}}, \, \alpha_4^{1^{uu}1^{dd}0^{ud}0^{ds}}.
\end{eqnarray}

If we consider the coupled equations corresponded to the reduced amplitudes
with two $s$-quarks ($ ^4_{\Lambda\Lambda} H$, $ ^4_{\Lambda\Sigma^0} H$,
$ ^4_{\Sigma^0\Sigma^0} H$), we use 33 equations (Tables \ref{tab3} -- \ref{tab5}).

The reduced amplitudes $\alpha_1$, $\alpha_2$ are equal to:

\begin{eqnarray}
\label{11}
&&\alpha_1^{1^{uu}}, \, \alpha_1^{1^{dd}}, \, \alpha_1^{0^{ud}}, \, \alpha_1^{0^{us}}, \, \alpha_1^{0^{ds}}, \, \alpha_1^{1^{ss}};
\end{eqnarray}

\begin{eqnarray}
\label{12}
&&\alpha_2^{1^{uu}1^{uu}}, \, \alpha_2^{1^{uu}1^{dd}}, \, \alpha_2^{1^{uu}0^{ud}},
 \, \alpha_2^{1^{uu}0^{us}}, \, \alpha_2^{1^{uu}0^{ds}}, \, \alpha_2^{1^{dd}1^{dd}},
 \nonumber
\\
&&\alpha_2^{1^{dd}0^{ud}}, \, \alpha_2^{1^{dd}0^{us}}, \, \alpha_2^{1^{dd}0^{ds}},
 \, \alpha_2^{0^{ud}0^{ud}}, \, \alpha_2^{0^{ud}0^{us}}, \, \alpha_2^{0^{ud}0^{ds}},
  \nonumber
\\
&&\alpha_2^{1^{uu}1^{ss}}, \, \alpha_2^{1^{dd}1^{ss}}, \, \alpha_2^{1^{ss}0^{ud}},
 \, \alpha_2^{0^{us}0^{us}}, \, \alpha_2^{0^{us}0^{ds}}, \, \alpha_2^{0^{ds}0^{ds}}.
\end{eqnarray}

The reduced amplitudes $\alpha_3$ and $\alpha_4$ are following:

\begin{eqnarray}
\label{13}
&&\alpha_3^{1^{uu}1^{dd}0^{ud}}, \, \alpha_3^{1^{uu}1^{dd}0^{us}}, \, \alpha_3^{1^{uu}1^{dd}0^{ds}};
\end{eqnarray}

\begin{eqnarray}
\label{14}
&&\alpha_4^{1^{uu}1^{dd}0^{ud}0^{ud}}, \, \alpha_4^{1^{uu}1^{dd}0^{ud}0^{us}}, \, \alpha_4^{1^{uu}1^{dd}0^{ud}0^{ds}},
  \nonumber
\\
&&\alpha_4^{1^{uu}1^{dd}0^{us}0^{us}}, \, \alpha_4^{1^{uu}1^{dd}0^{us}0^{ds}}, \, \alpha_4^{1^{uu}1^{dd}0^{ds}0^{ds}}.
\end{eqnarray}

It is interesting in the $ ^4_{\Omega} He$ hypernucleus. The system of equations
include the 38 equations (Table \ref{tab1}):

\begin{eqnarray}
\label{15}
&&\alpha_1^{1^{uu}}, \, \alpha_1^{1^{dd}}, \, \alpha_1^{0^{ud}}, \, \alpha_1^{0^{us}}, \, \alpha_1^{0^{ds}}, \, \alpha_1^{1^{ss}};
\end{eqnarray}

\begin{eqnarray}
\label{16}
&&\alpha_2^{1^{uu}1^{uu}}, \, \alpha_2^{1^{uu}1^{dd}}, \, \alpha_2^{1^{uu}0^{ud}},
 \, \alpha_2^{1^{uu}0^{us}}, \, \alpha_2^{1^{uu}0^{ds}}, \, \alpha_2^{1^{dd}1^{dd}},
 \nonumber
\\
&&\alpha_2^{1^{dd}0^{ud}}, \, \alpha_2^{1^{dd}0^{us}}, \, \alpha_2^{1^{dd}0^{ds}},
 \, \alpha_2^{0^{ud}0^{ud}}, \, \alpha_2^{0^{ud}0^{us}}, \, \alpha_2^{0^{ud}0^{ds}},
  \nonumber
\\
&&\alpha_2^{1^{uu}1^{ss}}, \, \alpha_2^{1^{dd}1^{ss}}, \, \alpha_2^{1^{ss}0^{ud}},
 \, \alpha_2^{0^{us}0^{us}}, \, \alpha_2^{0^{us}0^{ds}}, \, \alpha_2^{0^{ds}0^{ds}},
  \nonumber
\\
&&\alpha_2^{1^{ss}0^{us}}, \, \alpha_2^{1^{ss}0^{ds}};
\end{eqnarray}

\begin{eqnarray}
\label{17}
&&\alpha_3^{1^{uu}1^{uu}1^{dd}}, \, \alpha_3^{1^{uu}1^{uu}1^{ss}}, \, \alpha_3^{1^{uu}1^{dd}1^{ss}},
  \nonumber
\\
&&\alpha_3^{1^{uu}1^{uu}0^{ud}}, \, \alpha_3^{1^{uu}1^{dd}0^{ud}}, \, \alpha_3^{1^{uu}1^{ss}0^{ud}}, \, \alpha_3^{1^{dd}1^{ss}0^{ud}};
\end{eqnarray}

\begin{eqnarray}
\label{18}
&&\alpha_4^{1^{uu}1^{uu}1^{dd}1^{ss}}, \, \alpha_4^{1^{uu}1^{uu}1^{ss}0^{ud}}, \, \alpha_4^{1^{uu}1^{dd}1^{ss}0^{ud}},
 \nonumber
\\
&&\alpha_4^{1^{uu}1^{ss}0^{ud}0^{ud}}, \, \alpha_4^{1^{dd}1^{ss}0^{ud}0^{ud}}.
\end{eqnarray}

The analogous systems of equations are obtained for the other hypernuclei:
$ ^4_{\Sigma^+} He$ 31 equations, $ ^4_{\Sigma^-} He$ 31 equations,
$ ^4_{\Xi^0} He$ 41 equations, $ ^4_{\Xi^-} He$ 41 equations,
$ ^4_{\Lambda\Lambda} He$, $ ^4_{\Lambda\Sigma^0} He$, $ ^4_{\Sigma^0\Sigma^0} He$
33 equations, $ ^4_{\Sigma^+\Sigma^+} He$ 30 equations,
$ ^4_{\Sigma^-\Sigma^-} He$ 36 equations, $ ^4_{\Lambda\Sigma^+} He$ 35 equations,
$ ^4_{\Lambda\Sigma^-} He$ 40 equations, $ ^4_{\Sigma^+\Sigma^-} He$ 39 equations,
$ ^4_{\Sigma^+\Sigma^+} H$ 38 equations, $ ^4_{\Lambda\Sigma^+} H$ 42 equations,
$ ^4_{\Sigma^+\Sigma^-} H$ 44 equations.

The coefficients of the coupled equations are determined by the
permutation of quarks \cite{27, 28} (Appendix A). For simplicity, we
consider the graphical equations of the reduced amplitude
$\alpha_4^{1^{uu}1^{uu}1^{dd}0^{ds}}$ (See Fig. 1 Appendix B).
The interaction of quarks and gluons is constructed using the
functions $I_1$ -- $I_{46}$ (\ref{C1}) -- (\ref{C25}) in the Appendix C.

The main contributions are calculated using the functions $I_1$ and $I_2$:

\begin{eqnarray}
\label{31}
I_1(ij)&=&\frac{B_j(s_0^{13})}{B_i(s_0^{12})}
\int\limits_{(m_1+m_2)^2}^{\Lambda\frac{(m_1+m_2)^2}{4}}
\frac{ds'_{12}}{\pi}\frac{G_i^2(s_0^{12})\rho_i(s'_{12})}
{s'_{12}-s_0^{12}} \int\limits_{-1}^{+1} \frac{dz_1(1)}{2}
\frac{1}{1-B_j (s'_{13})}\, , \\
&&\nonumber\\
\label{32}
I_2(ijk)&=&\frac{B_j(s_0^{13}) B_k(s_0^{24})}{B_i(s_0^{12})}
\int\limits_{(m_1+m_2)^2}^{\Lambda\frac{(m_1+m_2)^2}{4}}
\frac{ds'_{12}}{\pi}\frac{G_i^2(s_0^{12})\rho_i(s'_{12})}
{s'_{12}-s_0^{12}}
\frac{1}{2\pi}\int\limits_{-1}^{+1}\frac{dz_1(2)}{2}
\int\limits_{-1}^{+1} \frac{dz_2(2)}{2}\nonumber\\
&&\nonumber\\
&\times&
\int\limits_{z_3(2)^-}^{z_3(2)^+} dz_3(2)
\frac{1}{\sqrt{1-z_1^2(2)-z_2^2(2)-z_3^2(2)+2z_1(2) z_2(2) z_3(2)}}
\nonumber\\
&&\nonumber\\
&\times& \frac{1}{1-B_j (s'_{13})} \frac{1}{1-B_k (s'_{24})}
 \, ,
\end{eqnarray}

\noindent
where $i$, $j$, $k$ correspond to the diquarks with the
spin-parity $J^P=0^+$, $1^+$.

We used the contributions of the functions (\ref{C1}) -- (\ref{C25}) in the Appendix C.
The other functions $I_i$ are small.

\section{Calculation results.}

The poles of the reduced amplitudes $\alpha_1$, $\alpha_2$, $\alpha_3$, $\alpha_4$
correspond to the bound states and determine the masses of low-lying
hypernuclei with the atomic number $A=4$ (Tables \ref{tab1} -- \ref{tab5}).

The experimental mass value of $ ^4_2 He$ is equal to $M=3727.4\, MeV$.
The experimental data of the hypernucleus $ ^4_{\Lambda} He$ $( ^4_{\Lambda} H)$
is $M=3922\, MeV$.

Our model uses only three parameters, which are determined by the following masses:
the cutoff $\Lambda=6.355$ and the gluon coupling constant $g=0.2122$.
The mass of the $u$, $d$ quarks is equal to $m=410\, MeV$, and the mass of strange quark
is $m_s=594\, MeV$, which takes into account the confinement potential (the shift
mass is equal to $37\, MeV$)

We used (as input) the hypernucleus $ ^4_{\Lambda} He$ $(ppn\Lambda)$. This
mass is equal to $M=3922\, MeV$ with the binding energy $B=10\, MeV$.
The similar equations give rise to the $ ^4_{\Lambda} H$ $(pnn\Lambda)$
with the mass $M=3922\, MeV$ and the binding energy $B=12\, MeV$
(Tables \ref{tab1}, \ref{tab2}). In the case of the hypernuclei $ ^4_{\Xi^0} H$ and
$ ^4_{\Xi^-} He$ with the mass $M=4121\, MeV$ the binding energy $B=12\, MeV$ and
$B=17\, MeV$ are calculated. The interesting results we obtained for the states
$ ^4_{\Xi^0} He$ and $ ^4_{\Xi^-} H$ with the mass $M=4106\, MeV$. The binding energies
are equal to $25\, MeV$ and $34\, MeV$, respectively.

The three hypernuclei $ ^4_{\Lambda\Lambda} H$, $ ^4_{\Lambda\Lambda} He$ and
$nn\Lambda\Lambda$ possess the negative binding energies $B_1=-8\, MeV$,
$B_2=-19\, MeV$ and $B_3=-15\, MeV$, respectively (Tables \ref{tab3} -- \ref{tab5}).
These state masses $M_1=4118\, MeV$, $M_2=4127\, MeV$ and $M_3=4127\, MeV$
are located near the energy thresholds. For the other nuclei the binding energies
are large, $\sim 50-150\, MeV$ (Tables \ref{tab1} -- \ref{tab5}).

We predict the eight hypernuclei with one $s$-quark. These states are determined
by the isospin $I=\frac{1}{2}$, $\frac{3}{2}$, $\frac{5}{2}$ and the spin-parity
$J^P=0^+$, $1^+$. Then the 38 hypernuclei with two $s$-quarks are calculated.
We obtained the two heavy hypernuclei $ ^4_{\Omega} H$ $(pnn\Omega)$ and
$ ^4_{\Omega} He$ $(ppn\Omega)$ with the mass equal to $M=4293\, MeV$
and the binding energies $B_1=198\, MeV$ and $B_2=196\, MeV$, respectively.
We derived only 16 different system equations, therefore the masses of
hypernuclei with $A=4$ are degenerated. The electromagnetic effect
is not included. The estimation of theoretical error is equal to
$1\, MeV$. This result has been obtained with the choice of the parameters
of the model. The spectrum of nuclei and hypernuclei changes from light
quark masses. In the recent work \cite{31} on the H-dibaryon and
nucleon-nucleon scattering length one shown this for even larger systems.

\section{Conclusions.}

The binding energies of a range of nuclei and hypernuclei with the atomic
number $A=4$ and strangeness $S\le 2$, including $ ^4 He$,
$ ^4_{\Lambda} He$ and $ ^4_{\Lambda\Lambda} He$ are calculated in
the limit of flavor $SU(3)$ symmetry at the physical strange quark
mass with quantum chromodynamics (without electromagnetic interactions)
\cite{15}. Infinite volume binding energies $B^{\infty} ( ^4_2 He)$,
$B^{\infty} ( ^4_{\Lambda} He)$, $B^{\infty} ( ^4_{\Lambda\Lambda} He)$
are given in Table \ref{tab6}.

In present paper the hypernucleus $ ^4_{\Lambda\Lambda} He$ with the mass
$M=4118\, MeV$ and the binding energy $B_1=-8\, MeV$ there is near
threshold. This state is similar to the multiquark systems with the small
decay width: the tetraquarks, pentaquarks and hexaquarks \cite{32}.
The experimental results on this state are not definite. The interactions
of nucleous in the nuclei can be considered using the phenomenological
nucleon-nucleon potential, which will be able to construct in the first
principles QCD. We believe that the nucleon-nucleon potentials describe
only small part of the quark-gluon interactions in the hadrons.
In this case the new approach will be able to consider the low-lying
hypernuclei and will allows us to construct the nonnucleon systems.

The all baryon using the strange quarks have the short life time.
In the nucleon medium their life time do not change essentially.
Therefore, the systems of 6, 9, 12 quarks are shortliving.
These results was discovered by the experimental data.
In the case of hadrons we will able to shown that the
rescattering hadron to be forbidden space corresponds,
to the interaction with the massive particles, which consist
of some quarks. The meson exchange model allows us to
obtain the only reduced picture of baryon interactions.
The new approach are constructed using the quark-gluon
structure of hypernuclei.

\begin{acknowledgments}
The authors would like to thank T. Barnes and L.V. Krasnov for useful discussions.
The reported study was partially supported by RFBR, research project No. 13-02-91154.
\end{acknowledgments}

\newpage

\appendix

\section{}

For instance, we consider the reduced amplitudes in the Appendix A: $\alpha_1^{1^{uu}}$ (page 10),
$\alpha_2^{1^{uu}1^{uu}}$ (page 11), $\alpha_3^{1^{uu}1^{uu}1^{dd}}$ (page 12) and
$\alpha_4^{1^{uu}1^{uu}1^{dd}0^{ds}}$ (page 13).

In Fig. 1 the coefficient of the term $I_{29}(1^{uu}1^{uu}1^{dd}0^{ds}1^{uu})\, \alpha_1^{1^{uu}}$
(page 10) is equal to 8, that is the number $8=2$ (the permutation of particles 1 and 2) $\times 2$
(the permutation of pairs (12) and (34)) $\times 2$ (we can replace the 9-th $u$-quark
with the 10-th $u$-quark); the coefficient of the term $I_{30}(1^{uu}1^{uu}1^{dd}0^{ds}1^{uu})\, \alpha_1^{1^{uu}}$
(page 10) is equal to 4, that is the number $4=2$ (the permutation of particles 1 and 2) $\times 2$
(the permutation of particles 3 and 4); the coefficient of the term
$I_{31}(1^{uu}1^{uu}1^{dd}0^{ds}1^{uu}1^{uu})\, \alpha_2^{1^{uu}1^{uu}}$
(page 11) is equal to 4, that is the number $4=2$ (the permutation of of pairs (12) and (78))
$\times 2$ (we can replace the 9-th $u$-quark with the 10-th $u$-quark); the coefficient of the term
$I_{32}(1^{uu}1^{uu}1^{dd}0^{ds}1^{uu}1^{uu})\, \alpha_2^{1^{uu}1^{uu}}$
(page 11) is equal to 8, that is the number $8=2$ (the permutation of particles 1 and 2)
$\times 2$ (the permutation of particles 3 and 4) $\times 2$ (we can replace the 9-th $u$-quark with
the 10-th $u$-quark); the coefficient of the term
$I_{37}(1^{dd}1^{uu}1^{uu}0^{ds}1^{dd}1^{uu}1^{uu})\, \alpha_3^{1^{uu}1^{uu}1^{dd}}$ (page 12)
is equal to 16, that is the number $16=2$ (the permutation of pairs (34) and (56)) $\times 2$
(we can replace the 10-th $u$-quark with the 11-th $u$-quark) $\times 2$
(the permutation of particles 1 and 2) $\times 2$ (we can replace the 9-th $d$-quark with
the 12-th $d$-quark); the coefficient of the term
$I_{38}(1^{uu}1^{uu}1^{dd}0^{ds}1^{uu}1^{uu}0^{ud})\, \alpha_3^{1^{uu}1^{uu}0^{ud}}$ (page 12)
is equal to 32, that is the number $32=2$ (the permutation of particles 1 and 2) $\times 2$
(the permutation of particles 3 and 4) $\times 2$ (the permutation of pairs (12) and (34))
$\times 2$ (we can replace the 9-th $u$-quark with the 11-th $u$-quark)
$\times 2$ (we can replace the 10-th $d$-quark with the 12-th $d$-quark); the coefficient of the term
$I_{46}(1^{uu}0^{ds}1^{uu}1^{dd}1^{uu}1^{uu}1^{dd}0^{ds})\, \alpha_4^{1^{uu}1^{uu}1^{dd}0^{ds}}$ (page 13)
is equal to 8, that is the number $8=2$ (the permutation of pairs (12) and (56)) $\times 2$
(we can replace the 9-th $u$-quark with the 10-th $u$-quark)
$\times 2$ (we can replace the 11-th $d$-quark with the 12-th $d$-quark).

The similar approach allows us to take into account the coefficients
in all the diagrams and equations.

\newpage

\section{}

\vskip60pt



\vskip60pt

Fig. 1. The graphical equations of the reduced amplitude $\alpha_4^{1^{uu}1^{uu}1^{dd}0^{ds}}$
$ ^4_{\Lambda} He$ $ppn\Lambda$ with the isospin projection $I_3=\frac{1}{2}$ and
the spin-parity $J^P=0^+$ (Table \ref{tab1}).

\newpage

\section{Some useful formulae.}

We used the functions
$I_1$, $I_2$, $I_3$, $I_4$, $I_5$, $I_6$, $I_7$, $I_8$, $I_9$, $I_{11}$, $I_{12}$, $I_{13}$,
$I_{14}$, $I_{15}$, $I_{16}$, $I_{18}$, $I_{23}$, $I_{24}$, $I_{29}$, $I_{30}$, $I_{31}$,
$I_{32}$, $I_{37}$, $I_{38}$, $I_{46}$:

\begin{eqnarray}
\label{C1}
I_1(ij)&=&\frac{B_j(s_0^{13})}{B_i(s_0^{12})}
\int\limits_{(m_1+m_2)^2}^{\frac{(m_1+m_2)^2\Lambda_i}{4}}
\frac{ds'_{12}}{\pi}\frac{G_i^2(s_0^{12})\rho_i(s'_{12})}
{s'_{12}-s_0^{12}} \int\limits_{-1}^{+1} \frac{dz_1(1)}{2}
\frac{1}{1-B_j (s'_{13})}\, , \\
&&\nonumber\\
\label{C2}
I_2(ijk)&=&\frac{B_j(s_0^{13}) B_k(s_0^{24})}{B_i(s_0^{12})}
\int\limits_{(m_1+m_2)^2}^{\frac{(m_1+m_2)^2\Lambda_i}{4}}
\frac{ds'_{12}}{\pi}\frac{G_i^2(s_0^{12})\rho_i(s'_{12})}
{s'_{12}-s_0^{12}}
\frac{1}{2\pi}\int\limits_{-1}^{+1}\frac{dz_1(2)}{2}
\int\limits_{-1}^{+1} \frac{dz_2(2)}{2}\nonumber\\
&&\nonumber\\
&\times&
\int\limits_{z_3(2)^-}^{z_3(2)^+} dz_3(2)
\frac{1}{\sqrt{1-z_1^2(2)-z_2^2(2)-z_3^2(2)+2z_1(2) z_2(2) z_3(2)}}
\nonumber\\
&&\nonumber\\
&\times& \frac{1}{1-B_j (s'_{13})} \frac{1}{1-B_k (s'_{24})}
 \, , \\
&&\nonumber\\
\label{C3}
I_3(ijk)&=&\frac{B_k(s_0^{23})}{B_i(s_0^{12}) B_j(s_0^{34})}
\int\limits_{(m_1+m_2)^2}^{\frac{(m_1+m_2)^2\Lambda_i}{4}}
\frac{ds'_{12}}{\pi}\frac{G_i^2(s_0^{12})\rho_i(s'_{12})}
{s'_{12}-s_0^{12}}\nonumber\\
&&\nonumber\\
&\times&\int\limits_{(m_3+m_4)^2}^{\frac{(m_3+m_4)^2\Lambda_j}{4}}
\frac{ds'_{34}}{\pi}\frac{G_j^2(s_0^{34})\rho_j(s'_{34})}
{s'_{34}-s_0^{34}}
\int\limits_{-1}^{+1} \frac{dz_1(3)}{2} \int\limits_{-1}^{+1}
\frac{dz_2(3)}{2} \frac{1}{1-B_k (s'_{23})} \, , \\
&&\nonumber\\
\label{C4}
I_4(ijk)&=&I_1(ik) \, , \\
&&\nonumber\\
\label{C5}
I_5(ijkl)&=&I_2(ikl) \, , \\
&&\nonumber\\
\label{C6}
I_6(ijkl)&=&I_1(ik) \times I_1(jl)
 \, , \\
&&\nonumber\\
\label{C7}
I_7(ijkl)&=&\frac{B_k(s_0^{23})B_l(s_0^{45})}{B_i(s_0^{12}) B_j(s_0^{34})}
\int\limits_{(m_1+m_2)^2}^{\frac{(m_1+m_2)^2\Lambda_i}{4}}
\frac{ds'_{12}}{\pi}\frac{G_i^2(s_0^{12})\rho_i(s'_{12})}
{s'_{12}-s_0^{12}}\nonumber\\
&&\nonumber\\
&\times&\int\limits_{(m_3+m_4)^2}^{\frac{(m_3+m_4)^2\Lambda_j}{4}}
\frac{ds'_{34}}{\pi}\frac{G_j^2(s_0^{34})\rho_j(s'_{34})}
{s'_{34}-s_{34}}
\frac{1}{2\pi}\int\limits_{-1}^{+1}\frac{dz_1(7)}{2}
\int\limits_{-1}^{+1} \frac{dz_2(7)}{2}
\int\limits_{-1}^{+1} \frac{dz_3(7)}{2}
\nonumber\\
&&\nonumber\\
&\times&
\int\limits_{z_4(7)^-}^{z_4(7)^+} dz_4(7)
\frac{1}{\sqrt{1-z_1^2(7)-z_3^2(7)-z_4^2(7)+2z_1(7) z_3(7) z_4(7)}}
\nonumber\\
&&\nonumber\\
&\times& \frac{1}{1-B_k (s'_{23})} \frac{1}{1-B_l (s'_{45})}
 \, , \\
&&\nonumber\\
\label{C8}
I_8(ijklm)&=&\frac{B_k(s_0^{15})B_l(s_0^{23})B_m(s_0^{46})}
{B_i(s_0^{12}) B_j(s_0^{34})}
\int\limits_{(m_1+m_2)^2}^{\frac{(m_1+m_2)^2\Lambda_i}{4}}
\frac{ds'_{12}}{\pi}\frac{G_i^2(s_0^{12})\rho_i(s'_{12})}
{s'_{12}-s_0^{12}}\nonumber\\
&&\nonumber\\
&\times&\int\limits_{(m_3+m_4)^2}^{\frac{(m_3+m_4)^2\Lambda_j}{4}}
\frac{ds'_{34}}{\pi}\frac{G_j^2(s_0^{34})\rho_j(s'_{34})}
{s'_{34}-s_0^{34}}\nonumber\\
&&\nonumber\\
&\times&\frac{1}{(2\pi)^2}\int\limits_{-1}^{+1}\frac{dz_1(8)}{2}
\int\limits_{-1}^{+1} \frac{dz_2(8)}{2}
\int\limits_{-1}^{+1} \frac{dz_3(8)}{2}
\int\limits_{z_4(8)^-}^{z_4(8)^+} dz_4(8)
\int\limits_{-1}^{+1} \frac{dz_5(8)}{2}
\int\limits_{z_6(8)^-}^{z_6(8)^+} dz_6(8)
\nonumber\\
&&\nonumber\\
&\times&
\frac{1}{\sqrt{1-z_1^2(8)-z_3^2(8)-z_4^2(8)+2z_1(8) z_3(8) z_4(8)}}
\nonumber\\
&&\nonumber\\
&\times&
\frac{1}{\sqrt{1-z_2^2(8)-z_5^2(8)-z_6^2(8)+2z_2(8) z_5(8) z_6(8)}}
\nonumber\\
&&\nonumber\\
&\times& \frac{1}{1-B_k (s'_{15})} \frac{1}{1-B_l (s'_{23})}
\frac{1}{1-B_m (s'_{46})}
 \, , \\
&&\nonumber\\
\label{C9}
I_9(ijkl)&=&I_3(ijl)
 \, , \\
&&\nonumber\\
\label{C10}
I_{11}(ijklm)&=&I_1(ik) \times I_2(jlm)
 \, , \\
&&\nonumber\\
\label{C11}
I_{12}(ijkl)&=&I_1(il)
 \, , \\
&&\nonumber\\
\label{C12}
I_{13}(ijklm)&=&I_2(ilm)
 \, , \\
&&\nonumber\\
\label{C13}
I_{14}(ijklm)&=&I_1(il) \times I_1(jm)
 \, , \\
&&\nonumber\\
\label{C14}
I_{15}(ijklm)&=&I_7(ijlm)
 \, , \\
&&\nonumber\\
\label{C15}
I_{16}(ijklmn)&=&I_8(ijlmn)
 \, , \\
&&\nonumber\\
\label{C16}
I_{18}(ijklmn)&=&I_1(il) \times I_2(jmn)
 \, , \\
&&\nonumber\\
\label{C17}
I_{23}(ijklmn)&=&I_2(ikl) \times I_2(jmn)
 \, , \\
&&\nonumber\\
\label{C18}
I_{24}(ijklmnp)&=&I_2(ilm) \times I_2(jnp)
 \, , \\
&&\nonumber\\
\label{C19}
I_{29}(ijklm)&=&I_1(im)
 \, , \\
&&\nonumber\\
\label{C20}
I_{30}(ijklm)&=&I_3(ijm)
 \, , \\
&&\nonumber\\
\label{C21}
I_{31}(ijklmn)&=&I_2(imn)
 \, , \\
&&\nonumber\\
\label{C22}
I_{32}(ijklmn)&=&I_1(im) \times I_1(jn)
 \, , \\
&&\nonumber\\
\label{C23}
I_{37}(ijklmnp)&=&I_1(im) \times I_2(jnp)
 \, , \\
&&\nonumber\\
\label{C24}
I_{38}(ijklmnp)&=&I_8(ijmnp)
 \, , \\
&&\nonumber\\
\label{C25}
I_{46}(ijklmnpq)&=&I_2(imn) \times I_2(jpq)
 \, .
\end{eqnarray}

\noindent
Here $i$, $j$, $k$, $l$, $m$, $n$, $p$, $q$ correspond to the diquarks with the
spin-parity $J^P=0^+, 1^+$.

The other functions $I_i$ can be neglected. The contributions of these functions
are smaller of few orders as compared the functions (\ref{C1}) -- (\ref{C25}). We do not take into account
these functions in the systems of coupled equations.


\newpage


\begin{thebibliography}{99}

\bibitem{1}
M. Fukugita, Y. Kuramashi, H. Mino, M. Okawa, and A. Ukawa,
Phys. Rev. Lett. {\bf 73}, 2176 (1994).

\bibitem{2}
M. Fukugita, Y. Kuramashi, M. Okawa, H. Mino, and A. Ukawa,
Phys. Rev. D{\bf 52}, 3003 (1995).

\bibitem{3}
S.R. Beane, P.F. Bedaque, K. Orginos, and M.J. Savage,
Phys. Rev. Lett. {\bf 97}, 012001 (2006).

\bibitem{4}
S.R. Beane, W. Detmold, H.-W. Lin, T.C. Luu, K. Orginos, M.J. Savage, A. Torok,
and A. Walker-Loud (NPLQCD Collaboration), Phys. Rev. D{\bf 81}, 054505 (2010).

\bibitem{5}
N. Ishii, S. Aoki, and T. Hatsuda, Phys. Rev. Lett. {\bf 99}, 002001 (2007).

\bibitem{6}
S. Aoki, T. Hatsuda, and N. Ishii, Comput. Sci. Dis. {\bf 1}, 015009 (2008).

\bibitem{7}
S. Aoki, T. Hatsuda, and N. Ishii, Prog. Theor. Phys. {\bf 123}, 89 (2010).

\bibitem{8}
T. Yamazaki, Y. Kuramashi, and A. Ukawa, Phys. Rev. D{\bf 84}, 054506 (2011).

\bibitem{9}
T. Yamazaki, Y. Kuramashi, and A. Ukawa, Phys. Rev. D{\bf 81}, 111504 (2010).

\bibitem{10}
P. de Forcrand and M. Fromm, Phys. Rev. Lett. {\bf 104}, 112005 (2010).

\bibitem{11}
S.R. Beane, E. Chang,  W. Detmold, H.-W. Lin, T.C. Luu, K. Orginos, A. Parreno, M.J. Savage, A. Torok,
and A. Walker-Loud, (NPLQCD Collaboration), Phys. Rev. D{\bf 85}, 054511 (2012).

\bibitem{12}
T. Inoue S. Aoki, T. Doi, T. Hatsuda, T. Ikeda, N. Ishii, K. Murano,
H. Nemura, and K. Sasaki (HAL QCD Collaboration), Nucl. Phys. A{\bf 881}, 28 (2012).

\bibitem{13}
S.R. Beane, E. Chang, S.D. Cohen, W. Detwold, H.W. Liu, I.C. Luu,
K. Orgince, A. Parreno, M.J. Savage, and A. Walker-Loud, arXiv: 1206.5219.

\bibitem{14}
T. Yamazaki, K.I. Ishikawa, Y. Kuramashi, and A. Ukawa,
Phys. Rev. D{\bf 86}, 074514 (2012).

\bibitem{15}
S.R. Beane, E. Chang, S.D. Cohen, W. Detwold, H.W. Liu, I.C. Luu,
K. Orgince, A. Parreno, M.J. Savage, and A. Walker-Loud,
(NPLQCD Collaboration), Phys. Rev. D{\bf 87}, 034506 (2013).

\bibitem{16}
S.M. Gerasyuta and E.E. Matskevich, Phys. Rev. D{\bf 82}, 056002 (2010).

\bibitem{17}
S.M. Gerasyuta, Z. Phys. C{\bf 60}, 683 (1993).

\bibitem{18}
S.M. Gerasyuta and E.E. Matskevich, arXiv:1211.0667.

\bibitem{19}
S.M. Gerasyuta and E.E. Matskevich, Phys. Rev. D{\bf 87}, 116006 (2013).

\bibitem{20}
J. Pochodzalla, Acta Phys. Pol. B{\bf 42}, 833 (2011).

\bibitem{21}
V.B. Kopeliovich, arXiv: 1203.3979.

\bibitem{22}
V.B. Kopeliovich, Nucl. Phys. A{\bf 721}, 1007 (2005).

\bibitem{23}
M. Agnello et al. [FINUDA Collaboration], Phys. Rev. Lett. {\bf 108}, 042501 (2012).

\bibitem{24}
G.'t Hooft, Nucl. Phys. B{\bf 72}, 461 (1974).

\bibitem{25}
G. Veneziano, Nucl. Phys. B{\bf 117}, 519 (1976).

\bibitem{26}
E. Witten, Nucl. Phys. B{\bf 160}, 57 (1979).

\bibitem{27}
O.A. Yakubovsky, Sov. J. Nucl. Phys. {\bf 5}, 1312 (1967).

\bibitem{28}
S.P. Merkuriev and L.D. Faddeev, Quantum Scattering Theory for System
of Few Particles (Nauka, Moscow, 1985) p. 398.

\bibitem{29}
V.V. Anisovich, S.M. Gerasyuta, and A.V. Sarantsev,
Int. J. Mod. Phys. A{\bf 6}, 625 (1991).

\bibitem{30}
G. Chew, S. Mandelstam, Phys. Rev. {\bf 119}, 467 (1960).

\bibitem{31}
Z. Davondi and M.J. Savage, Phys. Rev. D{\bf 84}, 114502 (2011).

\bibitem{32}
S.M. Gerasyuta, V.I. Kochkin, and E.E. Matskevich. Tetraquarks, pentaquarks, hexaquarks
in relativistic quark model (St. Petersburg University, St. Petersburg, 2012), p. 200.


\end{thebibliography}
\end{document}